\documentclass[12pt]{article}

\usepackage{epsfig,amsfonts,newlfont,rotate,float}
\begin{document}
\newcommand{\bbox}[1]{\mbox{\boldmath $#1$}}

\title{A model for the doped copper oxide compounds}

\author{J.L. Alonso,$^1$ Ph.\ Boucaud,$^2$ V. Mart\'\i n-Mayor$^3$\\
        and A.J. van der Sijs$^4$}

\date{\today}
\maketitle

\noindent {\small {\it $^1$Departamento de F\'\i sica Te\'orica, Universidad de 
Zaragoza,
50009 Zaragoza, Spain.}}

\noindent {\small {\it $^2$LPTHE, Universit\'e de Paris XI,
91405 Orsay Cedex, France.}}

\noindent {\small {\it $^3$Departamento de F\'{\i}sica Te\'orica I,
Universidad Complutense de Madrid,}} 

\noindent {\small {\it 28040 Madrid, Spain.}}

\noindent {\small {\it $^4$Swiss Center for Scientific Computing, ETH-Z\"urich,
ETH-Zentrum,}}

\noindent {\small {\it CH-8092 Z\"urich, Switzerland.}

\maketitle
\vfill

{\large
Short Title: A model for the doped copper oxides}

\begin{abstract}
We present a relativistic spin-fermion model for the cuprates, in
which both the charge and spin degrees of freedom are treated dynamically.
The spin-charge coupling parameter is associated with the doping fraction.
The model is able to
account for the various phases of the cuprates and their properties,
not only at low and intermediate doping but also for (highly)
over-doped compounds.
In particular, we acquire a qualitative understanding of
high-$T_c$ superconductivity through Bose-Einstein condensation of bound
charge pairs. The mechanism that binds these pairs 
does not require a Fermi sea.
\end{abstract}




\newpage

Ever since the discovery of high-$T_c$ superconductivity 
\cite{hightc} in the perovskites, with their characteristic
copper-oxide planes, there has been a great activity
in experimental work on these materials.
An outstanding feature is the presence of two-dimensional
antiferromagnetic correlations, which
persist in the superconducting phase at intermediate doping~\cite{Birgeneau}.
The absence of an isotope effect and the evidence for a
$d_{x^2-y^2}$-wave pairing state~\cite{Tsuei}, combined with the charge-2 nature
of the superconductivity, favour a mechanism based on spin rather than
phonon-mediated pair formation \cite{Schrieffer1,SchriefferAnderson}.

Here we present a simple model which is able to explain, at least
qualitatively, many of the observed properties of these materials,
from undoped to (highly) overdoped compounds.
We propose a relativistic, lattice-regularized, field-theoretical model of
interacting spins and fermions in 2+1 dimensions, with only two free parameters
in addition to the temperature:  a nearest-neighbour spin coupling $k$ and
a spin-fermion coupling $y$.
This model has much in common with microscopic spin-fermion
Hamiltonians \cite{mura}, which are somewhat less restrictive than
the $t$-$J$ model \cite{zhangrice}.

We consider the $(2+1)$-dimensional lattice field theory with action
\begin{eqnarray}
S \ =\ -k \sum_{n,\mu} \bbox{\vec\phi}_n\cdot
 \bbox{\vec\phi}_{n+\hat\mu}  \ + \ 
 \sum_{n,\mu} \frac{\rho}{2}\,  \bar\psi_n \gamma^\mu (\psi_{n+\hat\mu} -
    \psi_{n-\hat\mu}) 
\ +\ \lambda \sum_{n} \bar\psi_n \, \bbox{\vec\phi}_n \cdot
 \bbox{\vec\tau} \, \psi_n
\label{S}
    \, ,
\end{eqnarray}
which depends only on the ratio $y=\lambda/\rho$, through a change in
the normalization of the fermion field.
Here $n$ runs over a $(2+1)$-dimensional cubic Euclidean 
space-time lattice.
(We keep in mind, however, that some coherence perpendicular to the CuO$_2$
planes is required to avoid problems with the Hohenberg-Mermin-Wagner theorem
at finite temperature, and for a Bose-Einstein condensate to form).
In the present letter, our mean-field and numerical studies
will be limited to the zero-temperature case, 
corresponding to infinite Euclidean time direction.
However, we will argue on general grounds what
happens when the temperature is increased.

The fields $\psi$ represent the doped fermionic charges.
Each $\psi_n$ is a four-spinor consisting of two ``flavours'' of two-component
Dirac spinors.
The two flavours represent the two spin components of the (3+1)-dimensional
electron or hole.
Both components are taken in the same irreducible spinor representation
and the $4\times 4$ matrices $\gamma^\mu$ have the form
$\gamma^\mu = {\mathrm{diag}} (\sigma^\mu,\sigma^\mu)$, with Pauli matrices
$\sigma^\mu$ $(\mu=1,2,3)$.
The kinetic term for the fermions is of the nearest-neighbour (hopping) form.

The three-component fields $\bbox{\vec\phi}$ denote the spins
located at the copper ions.  They are real bosonic variables, subject to
the constraint $|\bbox{\vec\phi}_n|^2=1$, as in the
O(3) non-linear $\sigma$-model.
Their kinetic term, a nearest-neighbour hopping interaction, is the
field-theoretic equivalent of a Heisenberg superexchange interaction.

The last term in Eq.\ (\ref{S}) describes the spin-charge interaction,
which is diagonal in Dirac space.
The Pauli matrices $\tau^a$ act in flavour space.

We expect the doping to favour the hopping of the charge carriers
(that is the $\rho$ term in Eq.\ (\ref{S})).
Thus, the undoped compound will be described by $y=\infty$, where the model
simplifies to the non-linear $\sigma$-model, and the large-$y$ regime
corresponds to the strongly-correlated small-$x$ region, with immobile,
localized carriers \cite{footnote}.

We have investigated the model in the mean-field (MF) approximation,
supplemented with Monte Carlo (MC) simulations to determine the phase diagram.
The MF calculations are based on the saddle-point approach,
and follow closely the methods described in detail in
Ref.\ \cite{class}.
The MC simulations use the standard Hybrid Monte Carlo (HMC) 
algorithm,
adapted to deal with the $|\bbox{\vec\phi}_n|^2=1$ constraint.
The algorithm can be applied to the model (1), with two identical
fermion four-spinors to ensure a positive fermionic determinant.
The numerical work in this paper amounts to 16 days of the 32-processor
parallel computer RTNN based in Zaragoza,
which is equivalent to about 1.4 years of Pentium$^{\mathrm{(R)}}$ Pro CPU time.
Further details of the MF, large-$y$ expansions, and MC computations are relegated to a
forthcoming paper.

Fig.\ \ref{fig1} shows the phase diagram of the model at zero temperature.
Notice the influence of the dynamical charge carriers on the ordering of
the spins.
A similar phase structure is found in four-dimensional chiral
Yukawa models for the electroweak sector of the Standard Model of
elementary particle interactions \cite{class,bocketal,plzar}.

For $y=0$ one has the O(3) non-linear $\sigma$-model, with its well-known
disordered, paramagnetic (PM) and ordered ferromagnetic (FM) and
antiferromagnetic (AFM) phases.
This field theory, in either of its ordered phases,
describes the nearest-neighbour AFM Heisenberg model,
as was shown rigorously in the $S\rightarrow\infty$ limit \cite{haldane},
and convincingly argued to hold also for finite spin $S$ \cite{chakraetal,HN}.
This Heisenberg model is
known to provide a very good quantitative description of the spin
background of the copper oxide layers in the undoped parent compounds
such as La$_2$CuO$_4$ \cite{exp}.

For $y>0$, the model contains
dynamical fermions, and the phases of the O(3) model extend into the $y$--$k$ plane.
At a certain value of $y$, the PM--FM and PM--AFM transition
lines meet in a point $A$ and the disordered phase ends.

In the limit $y\rightarrow\infty$, the fermions can be integrated out
and one {\em also\/} arrives at the O(3) non-linear sigma model.
For large but finite $y$ the three phases
extend into the $y$--$k$ plane again, with the PM--FM and PM--AFM
lines meeting in the point $B$.
At strong coupling, one would expect an evolution from
commensurate to incommensurate AFM ordering,
with increasing doping~\cite{SS}.
After integrating out the fermion fields in our model, 
frustrating couplings (of order $1/y^4$ in the strong coupling expansion)
are generated,
which presumably leads to such a phenomenon.
Our MC simulation was done
on a too small lattice to be able to resolve such an effect,
and our MF is not reliable in this intermediate $y$ region~\cite{class}.

In the intermediate-$y$ region
between $A$ and $B$, at negative $k$, there is a phase with both FM
and AFM properties.  Both the magnetization and the staggered
magnetization are found to be non-zero here, in the MC
simulation.  A canted state could account for such behaviour, for
example.  However, it is difficult to characterize this phase in a
precise way numerically, and we cannot at present exclude other less
conventional behaviour, such as an incommensurate (spiral) ground
state.  For now, we will simply label the order in this phase as
``exotic magnetism'' (EM).

Note that the PM phases in the small and
large-$y$ regions are not connected.  This crucial property, which
radically distinguishes our model from other models for the cuprates,
admits the possibility of qualitatively different behaviour in these
phases.  For this reason, these phases will be denoted as PMW and PMS
(W and S for weak and strong), respectively.  No phase transition was
found to separate the small and large-$y$ regions in the FM phase,
which we will call FM(W) and FM(S).  There may be a crossover, though.
Similarly, we use the terminology
AFM(W) and AFM(S) to distinguish the different regions in the AFM
phase, which are presumably connected below some large negative value
of $k$ (although we have not explored this region numerically).

Let us now consider the relevant excitations in this phase
diagram. At weak coupling, we have combined the mean-field
approximation with perturbation theory in $y$~\cite{class}.  
Qualitatively, the result can be understood  if 
$\bbox{\vec\phi}$ is replaced by a MF value in Eq.\ (1).
In the disordered
PMW phase one finds massless fermions, while in the FM(W) phase the
fermion mass $m$ goes like $yM$, where $M=|\langle \bbox{\vec\phi} \rangle|$
is the magnetization.  In the AFM(W) phase the situation is similar,
although a momentum-space diagonalization is required first, as we
will discuss for the AFM(S) case below.
This weak-coupling part of the phase diagram is characterized by
propagating fermions, {\em i.e.}, we will have Fermi-liquid behaviour.

In the large-$y$ region the situation is {\em very\/} different.
A change of variables is required for perturbation theory:
$\psi\rightarrow (1/y)\,
\bbox{\vec\phi} \cdot \bbox{\vec\tau}\, \psi$,
$\bar\psi\rightarrow \bar\psi$,
which brings 
$(1/y) \, \bbox{\vec\phi} \cdot \bbox{\vec\tau}$
into the kinetic term for the fermions \cite{class}.
In particular, in the PMS phase, as $\langle \bbox{\vec\phi} \rangle = 0$,
the kinetic term vanishes in the MF approximation, 
and the fermions are
essentially non-propagating, implying insulating behaviour at the MF level.

A crucial result we obtain is the dynamical generation of spin-zero, bosonic
bound states of fermions in the PMS phase.  In a MF calculation of the
two-point correlation function of the composite, spin-singlet field
$\varepsilon_{ab}\psi_x^a\psi_x^b$ ($a$, $b$ flavour indices), we find
for the mass $m_{\mathrm{PMS}}^2= 4(y^2-3/2)$, suggesting that this
excitation becomes light if one is sufficiently far away from
$y=\infty$.  We shall refer to these light charge-2 excitations as
``PMS pairs''.  A very similar example of bound state
formation from fermions strongly coupled to disordered scalars has
been found in a different context in Ref.\ \cite{PMSMF}. We emphasize that
this mechanism for forming charge pairs does not require a Fermi sea. 
As a final remark let us mention that one
cannot exclude in this region a rich spectrum of bound states with
different quantum numbers. We have started a detailed MC study of such
bound states.

In the AFM(S) phase, the AFM alignment of the spins gives rise
to a coupling between fermion momenta $p$ and $p+(\pi,\pi,\pi)$.
After momentum-space diagonalization, the denominator of the fermion
propagator has the form $y^2 - v^2\sum_\mu \sin^2(p_\mu)$
($v$ is the staggered magnetization), which is unusual for a Euclidean
space-time lattice.
We note, however, that this
expression suggests the possibility of light excitations with a
relativistic dispersion relation (with $m^2 = (y^2-3v^2)/v^2$ in this naive
MF approximation) around spatial momenta
$(\pm\pi/2,\pm\pi/2)$.  Hole pockets \cite{Schrieffer1,shraisig} thus
appear to emerge naturally as the only candidates for low-energy excitations
in this phase. Note that, in the context of our model, they are only
expected in this AFM phase, not at superconducting doping concentrations
which correspond to the PMS phase. For some recent experimental
results see Ref.\ \cite{NORMAN}.

Next we will analyse how our model describes the various phenomena in
the cuprates.
The undoped material is an AFM Heisenberg model which is described by a point
with $y=\infty$ and $k$ slightly smaller than $-k_c$, in the AFM(S) phase.  The actual value
of $k$ is related to the spin stiffness and velocity; in the large-$S$
approximation $k\propto S$ \cite{chakraetal}. 
Note that by virtue of the $k \rightarrow -k$ symmetry at $y=\infty$
we are free to take $k$ in the AFM phase (and as we shall see next,
this is in fact the only natural choice in the context of our model),
although the original papers \cite{haldane,chakraetal,HN} relating the
AFM Heisenberg model to the (continuum) O(3) model focused on its FM phase.
 As the doping fraction $x$
is increased, the carrier mobility increases which we argued to correspond
to decreasing $y$ in our model (cf.\ the line of arrows in Fig.\ \ref{fig1}).
One might assume
a relation of the type $x\sim C^2/(C^2+y^2)$ for some constant $C$ (in the
case $C=1$ the horizontal axis in Fig.\ \ref{fig1} would then correspond to
$1-x$), but this is immaterial for the qualitative picture.
Note that the downward tendency of the PMS--AFM(S) transition with decreasing
$y$ is consistent with the experimentally observed reduction of AFM order
upon doping.

At some point within the AFM(S) phase, the hole pockets could become light
and start to dominate.
When the doping is increased even more, we
move into the PMS phase.
Close to the transition, short-range AFM correlations are still present,
in agreement with experiment.
In this phase, the only light excitations are PMS pairs.
At temperature $T=0$ they will be Bose-Einstein (BE) condensed, leading to
superconductivity.
Since a finite number ${\cal O}(x_{c_1}/2)$ of them becomes available
at the critical doping $x_{c_1}$,
one expects a finite ({\em i.e.}, not infinitesimally small) critical
temperature $T_c$ here.
Following the arrow line towards
even smaller $y$, we leave the superconducting PMS phase at $x_{c_2}$.
Again, $T_c$ is expected to remain finite up to $x_{c_2}$.

For very large doping we reach the weak-coupling region with
Fermi-liquid behaviour.
We will end up in the PMW or the AFM(W) phase,
depending on the actual trajectory of the system with doping.
In either case, the effective mass of the carriers,
generated through the interaction with the spin waves, will be zero at
the MF level, as the masses are proportional to the magnetization here.
This illustrates how such widely varying behaviour in the cuprates, controlled
by the doping fraction $x$, 
is reproduced by varying just one parameter in our model.

In between, in the intermediate-$y$ region, our model predicts a FM phase.
The system traverses a region where a transition takes place from
a description of the charge carriers in terms of the variables
$(1/y)\,\bbox{\vec\phi} \cdot \bbox{\vec\tau} \, \psi$
at large $y$, to a description in terms of the fields
$\psi$ in the \hbox{small-$y$} Fermi-liquid regime.
There may be a kind of cross-over in this region.

What happens when the temperature is increased from zero?
So far, we do not have any rigorous MF or MC results at non-zero
temperature.  However, by the following fairly general arguments we are led
to conjecture Fig.~2 as a schematic sketch of the most general phase diagram
suggested by the model.
Let us take $x$ such that we are well inside the PMS phase,
with superconductivity at $T=0$.
At a certain temperature $T_c=T_{\mathrm {BE}}$, the BE condensation will be undone.
Above $T_c$ there is no superconductivity
but the fermions are still bound together in light PMS pairs,
which we expect to participate in ``normal'', charge-2 conduction.
This phase is sometimes called a ``spin-gap'' phase.
Above some spin-gap temperature, $T_{\mathrm {SG}}$, 
these PMS pairs
will be broken apart by thermal fluctuations. 
Here the only possible carriers would be fermions,
but as we have already discussed, their kinetic term vanishes in a 
strong-coupling paramagnetic phase, in the mean field approximation.
This perfectly insulating behaviour will not survive beyond the mean-field
approximation, though.

In Fig.~2, we have supposed that $T_{\mathrm {SG}}$ is always larger than
$T_{\mathrm {BE}}$. 
Depending on the variation of the mass and the binding energy of the
bound state with doping, it could happen that the $T_{\mathrm {SG}}$ and
$T_{\mathrm {BE}}$ critical lines join at some point. 
The actual situation has to be investigated numerically.

An interesting new prediction of our model is that
superconductivity is unlikely to occur in materials with $S>1/2$, as
this would correspond to $k \ll -k_c$ \cite{chakraetal},
making it much more difficult to pass through the PMS phase.
For instance, upon doping the layered compounds La$_{1-x}$Sr$_{1+x}$MnO$_4$, which have localized $S=3/2$ spins (or $k\sim -2$)
a disappearance of 
the antiferromagnetic phase and the subsequent emergence of 
an exotic magnetic phase (maybe a spin-glass phase) is observed
\cite{MORITOMO}, but no superconductivity appears.

Another intriguing observation is the possibility of re-entrant
superconductivity for certain values of $x$, 
with superconductivity restricted to a temperature interval
\hbox{$0<T_{M} < T < T_{c}$.}
For example, for
$x$ slightly smaller than $x_{c_1}$ we would be in the AFM(S) phase,
at a point $(y,k)$ very close to the transition to PMS.  Raising the
temperature now has the effect of increasing the magnetic disorder.  
Let us assume this disorder to be similar
to the disorder coming from the dynamics ({\em i.e.}, doping at zero
temperature). Then
the
PMS phase will be enlarged and absorb the point $(y,k)$, implying
superconductivity provided that we are still below $T_{\mathrm {BE}}$.
This argument requires $T_{\mathrm {BE}}$ to be non-zero at $x_{c_1}$, which is
plausible.
Similar re-entrant behaviour is expected for $x$ slightly larger than
$x_{c_2}$. Some experimental evidence for reentrance has been found some years
ago \cite{BREWER} in YBaCuO, but it seems to have gone largely unnoticed.
Further experiments are required to resolve this question.
Some knowledge of the mass and the binding energy of the PMS-pairs
is required in order to make quantitative predictions for this surprising
phenomenon.

In conclusion, we have formulated and investigated a simple, relativistic
spin-fermion field theory in 2+1 dimensions
capable of explaining, at least qualitatively, a variety of
experimental properties of the cuprate superconductors and their parent
compounds as a function of doping fraction and temperature.
In particular, our model provides a qualitative understanding of insulating
AFM behaviour at low doping, of high-$T_c$ superconductivity through the
Bose-Einstein condensation of spin-disorder-bound charge pairs at
intermediate doping, and of Fermi-liquid behaviour at large doping.
In addition we have formulated several predictions, which may be amenable
to experimental testing; presumably, more predictions can be derived.

We are indebted to A. Muramatsu for very helpful remarks and
stimulating discussions.
We also acknowledge interesting discussions with A.~Cruz, Ph.~de Forcrand,
J.M. De Teresa, J.G.~Esteve, D.~Frenkel, J.~Garcia, M.R. Ibarra, 
R.~Mahendiran, R.~Navarro,  C.~Rillo and especially G.~Sierra.
We thank L.A.~Fern\'andez, A.~Taranc\'on and the others members of the
RTNN collaboration for computing facilities.
This work is financially supported by DGICYT (Spain), project
AEN 96-1670, and by Acci\'on Integrada Hispano-Francesa HF1996-0022.

\newpage

\begin{figure}[htb]
\caption{Phase diagram at zero temperature, for
the model of Eq.\~(\protect\ref{S}).  Dashed lines are MF predictions,
solid lines with data points are from a MC simulation on an
$8^3$ lattice, with two families of dynamical fermions, as required by
the HMC method. The line of arrows illustrates the ``evolution''
of the material with increasing doping (see main text). The dotted vertical
line indicates a crossover.}
\label{fig1}
\end{figure}

\begin{figure}[htb]
\caption{Sketch of the predicted phase structure in the $x$-$T$ plane.
The dotted vertical line at $x_{\mathrm{S}-\mathrm{W}}$ indicates a 
crossover.}
\label{fig2}
\end{figure}

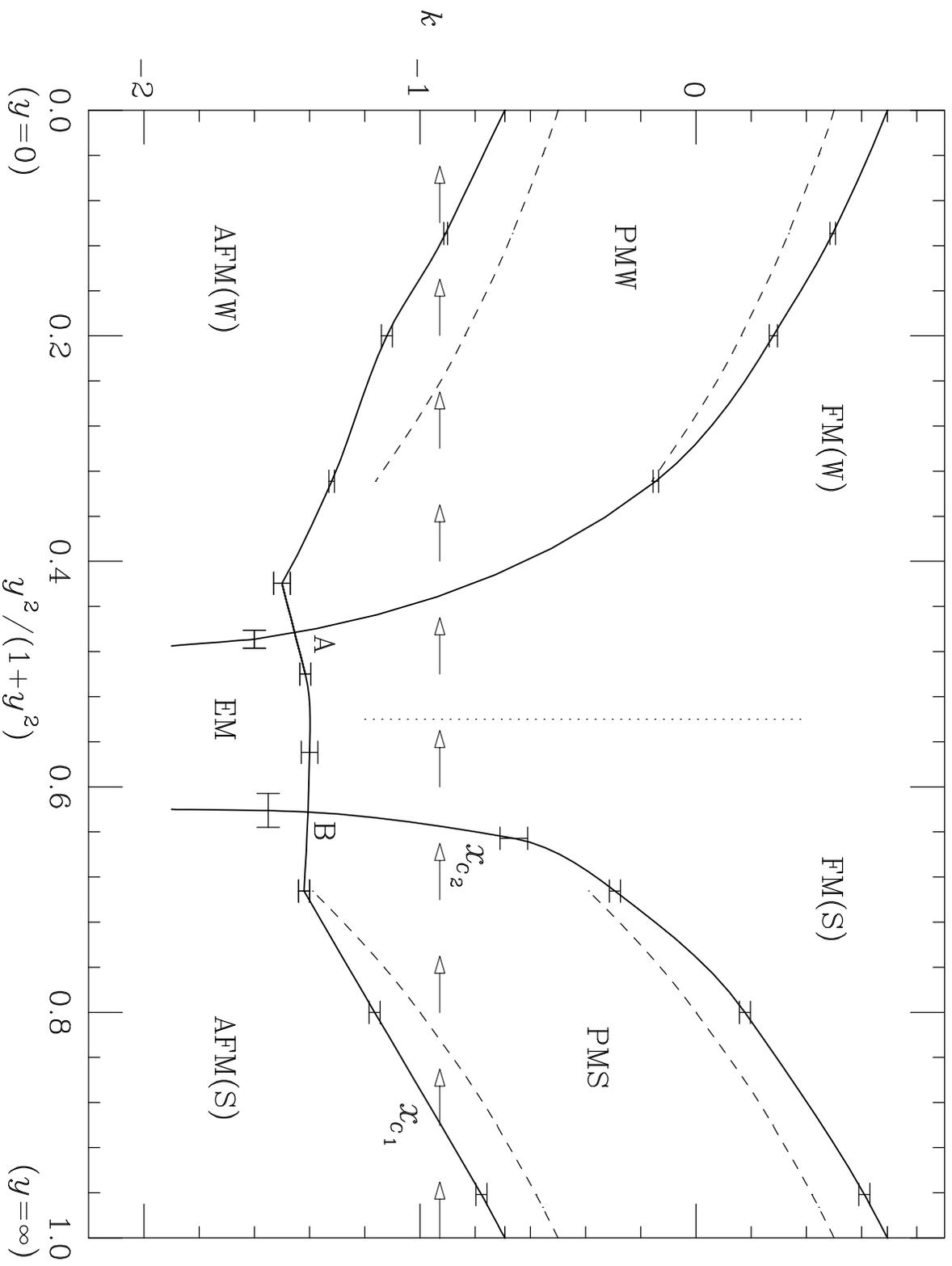
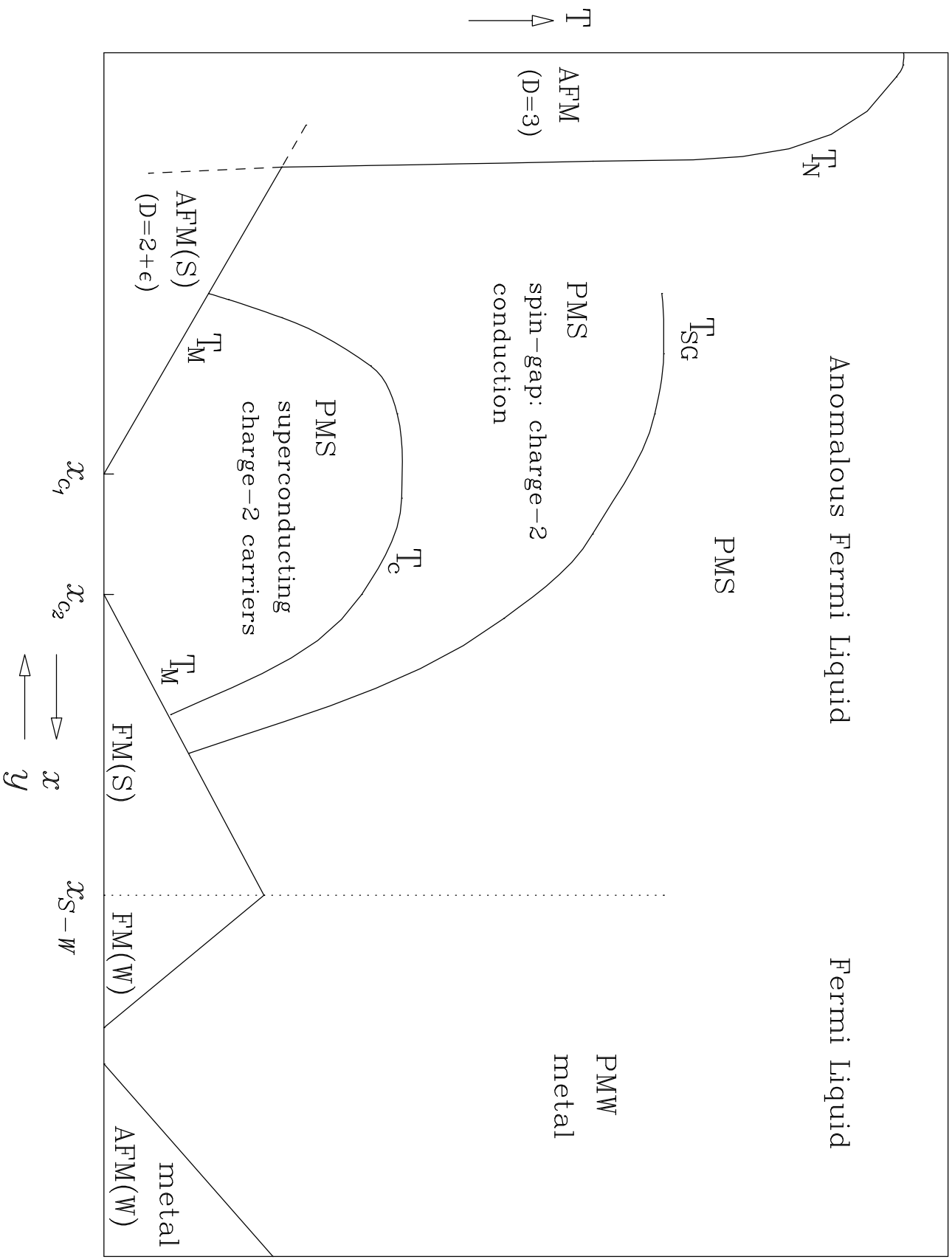
\end{document}